\documentclass[final,11pt]{elsarticle}

\usepackage{latexsym}

\usepackage{amssymb}
\usepackage{amsmath}

\usepackage{xcolor}

\usepackage{multirow}

\usepackage{adjustbox}

\usepackage{hyperref}

\journal{Journal of High Energy Astrophysics}

\hypersetup{linkcolor=cyan,citecolor=magenta,filecolor=yellow,urlcolor=blue}

\begin{document}

\begin{frontmatter}



\title{On the formation of strange quark stars from supernova in compact binaries}

\author[uchile,icranetpescara,icra]{L.~M.~Becerra}
\ead{laura.becerra@umayor.cl}

\author[bsc]{F.~Cipolletta}
\ead{federico.cipolletta@bsc.es}

\author[unife,infnferrara]{A.~Drago}
\ead{drago@fe.infn.it}

\author[unife,infnferrara]{M.~Guerrini}
\ead{mirco.guerrini@unife.it}

\author[polito,infntorino]{A.~Lavagno}
\ead{andrea.lavagno@polito.it}

\author[unife,infnferrara]{G.~Pagliara}
\ead{pagliara@fe.infn.it}

\author[icranetpescara,icra,icranetferrara,unife,inafrome]{J. A. Rueda} 
\ead{jorge.rueda@icra.it}


\affiliation[uchile]{organization={Centro Multidisciplinario de Física, Vicerrectoría de Investigación, Universidad Mayor},
            city={Santiago},
            postcode={8580745}, 
            country={Chile}}

\affiliation[bsc]{organization={Barcelona Supercomputing Center (BSC)},
            addressline={Plaça d'Eusebi Güell 1-3}, 
            city={Barcelona},
            postcode={08034}, 
            country={Spain}}

\affiliation[infnferrara]{organization={INFN, Sezione di Ferrara},
            addressline={Via Saragat 1}, 
            city={Ferrara},
            postcode={I-44122}, 
            country={Italy}}

\affiliation[polito]{organization={Dipartimento Scienza Applicata e Tecnologia, Politecnico di Torino},
            city={Torino},
            postcode={I-10129}, 
            country={Italy}}    

\affiliation[infntorino]{organization={INFN, Sezione di Torino},
            city={Torino},
            postcode={I-10125}, 
            country={Italy}}
            
\affiliation[icranetpescara]{organization={ICRANet},
            addressline={Piazza della Repubblica 10}, 
            city={Pescara},
            postcode={I-65122}, 
            country={Italy}}

\affiliation[icra]{organization={ICRA},
            addressline={Dipartimento di Fisica, Sapienza Università di Roma}, 
            city={Rome},
            postcode={I-00185}, 
            country={Italy}}

\affiliation[icranetferrara]{organization={ICRANet-Ferrara, Dipartimento di Fisica e Scienze della Terra, Università degli Studi di Ferrara},
            addressline={Via Saragat 1}, 
            city={Ferrara},
            postcode={I-44122}, 
            country={Italy}}

\affiliation[unife]{organization={Dipartimento di Fisica e Scienze della Terra, Università degli Studi di Ferrara},
            addressline={Via Saragat 1}, 
            city={Ferrara},
            postcode={I-44122}, 
            country={Italy}}

\affiliation[inafrome]{organization={INAF, Istituto de Astrofisica e Planetologia Spaziali},
            addressline={Via Fosso del Cavaliere 100}, 
            city={Rome},
            postcode={I-00136}, 
            country={Italy}}

\begin{abstract}
Strange quark stars (SQSs), namely compact stars entirely composed of deconfined quark matter, are characterized by similar masses and compactness to neutron stars (NSs) and have been theoretically proposed to exist in the Universe since the 1970s. However, multiwavelength observations of compact stars in the last 50 years have not yet led to an unambiguous SQS identification. This article explores whether SQSs could form in the supernova (SN) explosion of an evolved star (e.g., carbon-oxygen, or Wolf-Rayet) occurring in a binary with the companion being a neutron star (NS). The collapse of the iron core of the evolved star generates a newborn NS and the SN explosion. Part of the ejected matter accretes onto the NS companion as well as onto the newborn NS via matter fallback. The accretion occurs at hypercritical (highly super-Eddington) rates, transferring mass and angular momentum to the stars. We present numerical simulations of this scenario and demonstrate that the density increase in the NS interiors during the accretion process may induce quark matter deconfinement, suggesting the possibility of SQS formation. We discuss the astrophysical conditions under which such a transformation may occur and possible consequences.
\end{abstract}





\end{frontmatter}
%

\section{Introduction}\label{sec:1}

Compact stars can reach densities several times the nuclear saturation density in their core. Thus, they are one of the best candidates in the Universe where a phase of deconfined strange quark matter (SQM) can be found \cite{Annala:2019puf}. The phenomenology of compact stars strongly depends on the value of the (unknown) energy per baryon of $\beta$-stable SQM at vanishing pressure and temperature in bulk, denoted by $(E/N)_0$. If $(E/N)_0 >930$ MeV, namely larger than the energy per baryon of iron, then SQM and hadronic matter (HM) are separated by a phase transition line in the QCD phase diagram. At increasing density (and fixed temperature), one finds first HM, then, at a baryon density $n_{mp}$, a mixed phase of HM and SQM, and finally, at a baryon density $n_{pq}$, pure SQM. This implies that compact stars with a central baryonic density  $n_c > n_{mp}$ and below $n_{pq}$ would contain an HM layer and a core of mixed phase. Compact stars with $n_c>n_{pq}$ would contain an HM layer, a mixed phase layer, and a core of pure SQM. In both cases ($n_c>n_{mp}$ and $n_c>n_{pq}$), those objects are called hybrid stars (see, e.g., \cite{Alford:2004pf}). 
On the other hand, if $(E/N)_0 <930$ MeV, SQM is stable (the so-called Bodmer-Witten hypothesis, \cite{Bodmer:1971we,Witten:1984rs}). In this scenario, after a first seed of deconfined quarks is formed at a density $n_{\rm crit}$ in an NS, the entire matter of the object is converted to SQM \cite{Drago:2015fpa}. Thus, compact stars are neutron stars (NS) or strange quark stars (SQSs), the latter entirely composed of SQM.  This second possibility leads to the so-called two-family scenario, in which NSs and SQSs coexist \cite{Drago:2013fsa} (see also \cite{2003ApJ...586.1250B,2004ApJ...614..314B}), and it is the one we will assume here. The two-family scenario has a rich phenomenology, which has been investigated in several papers (see, e.g., \cite{Drago:2015cea,Drago:2015dea,DePietri:2019khb,Bombaci:2020vgw}). The main prediction of this scenario pertains to the structure of compact stars. While NSs could be very compact (with radii smaller than about $11$ km) but light (with maximum masses significantly below $2M_\odot$), SQSs could be very massive and large \cite{Bombaci:2020vgw}. In turn, this scenario resolves the so-called hyperon puzzle \cite{Lonardoni:2014bwa}: the necessary appearance of hyperons at high densities could soften the hadronic equation of state (EOS) too much, leading to maximum masses substantially smaller than $2M_\odot$, within the standard one-family scenario \cite{Baldo:1999rq}. In the two-families scenario, instead, the formation of hyperons triggers the conversion of the NS into a SQS whose EOS is stiff enough to support masses up to $2.6M_\odot$ \cite{Bombaci:2020vgw}. Another interesting feature within this scenario is the possibility of explaining the existence of very low mass stars, as the one possibly associated with HESS J1731--347 \cite{Doroshenko:2022nwp}, that could originate from the collapse of progenitors that have captured a seed of SQM from the galactic flux of strangelets \cite{DiClemente:2022wqp}.

A crucial issue that needs to be addressed is identifying which astrophysical phenomena lead to deconfinement (and thus to the formation of a SQS) and which signatures can be associated with them. A possibility has been suggested in \cite{Fischer:2017lag}: deconfined quark matter (QM) could form during the core-collapse supernovae (CCSNe) associated with blue supergiant stars, and, thanks to the enormous latent heat released by the hadron-quark phase transition, it can provide a mechanism for the explosions of such massive stars. Additionally, superluminous CCSNe (i.e., CCSNe with significantly brighter emission than ordinary ones) may originate from this mechanism \cite{Kuroda_2022}. Indeed, their typical explosion energies ($\sim 10^{51}$ erg \cite{Fiore:2021jyo}) and distinctive light-curve shapes (e.g., \cite{Inserra:2017uyc}) are not explained by standard neutrino-driven explosions (e.g., \cite{Umeda:2007wk}).

Another possible path is related to the mergers of two compact stars, in particular, NS-NS mergers. We recall that these mergers are considered as the primary progenitors of the short-duration gamma-ray bursts (GRBs, see, e.g., \cite{1986ApJ...308L..47G, 1986ApJ...308L..43P, 1989Natur.340..126E,1991ApJ...379L..17N}). In the two-families scenario, an NS-NS merger could lead to the formation of a hypermassive, supramassive, or even stable SQS (the formation of a NS might be unlikely due to the expected large mass of the remnant), depending on the total mass and asymmetry of the binary \cite{DePietri:2019khb}. Interestingly, the formation of SQSs or hybrid stars after the merger could be distinguished by their specific signatures within the post-merger gravitational wave signal (see, e.g, \cite{Bauswein:2015vxa,Prakash:2021wpz}, and references therein), which might be revealed by the next generation detectors.

In addition, an interesting suggestion is that QM deconfinement could arise in low-mass X-ray binaries \cite{Wiktorowicz:2017swq}: the material accreted onto the NS leads to a compression of matter and an increase of the central density up to the critical density $n_{\rm crit}$ for quark deconfinement to start. This suggestion leads us to explore the possibility of quark deconfinement in a late evolutionary stage of such systems in which accretion onto an NS can be more extreme: the second SN explosion that occurs in the binary. After the system survives the X-ray binary phase and possibly multiple common-envelope phases, it is formed by a compact binary composed of an evolved star and the NS formed in the first SN explosion in the binary evolution. These systems can include the ultra-stripped binaries that are considered likely progenitors of NS-NS binaries \cite{2015MNRAS.451.2123T}, as well as the binary progenitors of long-duration GRBs in the the binary-driven hypernova (BdHN) scenario, which predicts a variety of GRB features and consequent binary outcomes (NS-NS or NS-BH) depending on the binary parameters (see, e.g., \cite{2012ApJ...758L...7R, 2014ApJ...793L..36F,2015PhRvL.115w1102F,2015ApJ...812..100B,2016ApJ...833..107B,2019ApJ...871...14B,2022PhRvD.106h3004R,2022PhRvD.106h3002B,2023ApJ...955...93A,2024ApJ...976...80B,2025PhRvD.111b3010R}, and references therein). 

Therefore, in this work, we explore the possible occurrence of NS-SQS conversion under the massive accretion process that can occur in an SN exploding in a compact-orbit binary composed of an evolved star (e.g., CO, Wolf-Rayet, or similar) and a NS companion. If one or both of the NSs in the binary (the newborn NS out of the CO core-collapse and the NS companion) convert to SQSs due to the substantial amount of mass accreted, these systems can form as remnants of the cataclysmic event, NS-NS, SQS-SQS, or NS-SQS binaries. While the conversion of an NS into a SQS must necessarily be accompanied by a significant amount of energy release that can power energetic transient phenomena, we confine ourselves in this work to a first suite of numerical simulations that allow us to identify the plausible physical and astrophysical conditions for the occurrence of SQM deconfinement by the massive accretion in the final evolutionary stages of compact-orbit binaries. The analysis of how such a powerful energy release could be transformed into observable electromagnetic emission is left for future studies. 

The organization of this work is as follows: Sec. \ref{sec:2} provides an overview of accretion rates derived from numerical simulations. In Sec. \ref{sec:nsevolution}, we describe the setup used to model the time evolution of the NSs. Sec. \ref{sec:eos} introduces the EOS considered, as well as the criteria for converting a NS into a SQS. The results of our analysis are discussed in Sec. \ref{sec:results}, and we summarize our main findings in Sec. \ref{sec:conclusions}.

\begin{figure*}
    \centering
    \includegraphics[width=0.33\hsize,clip]{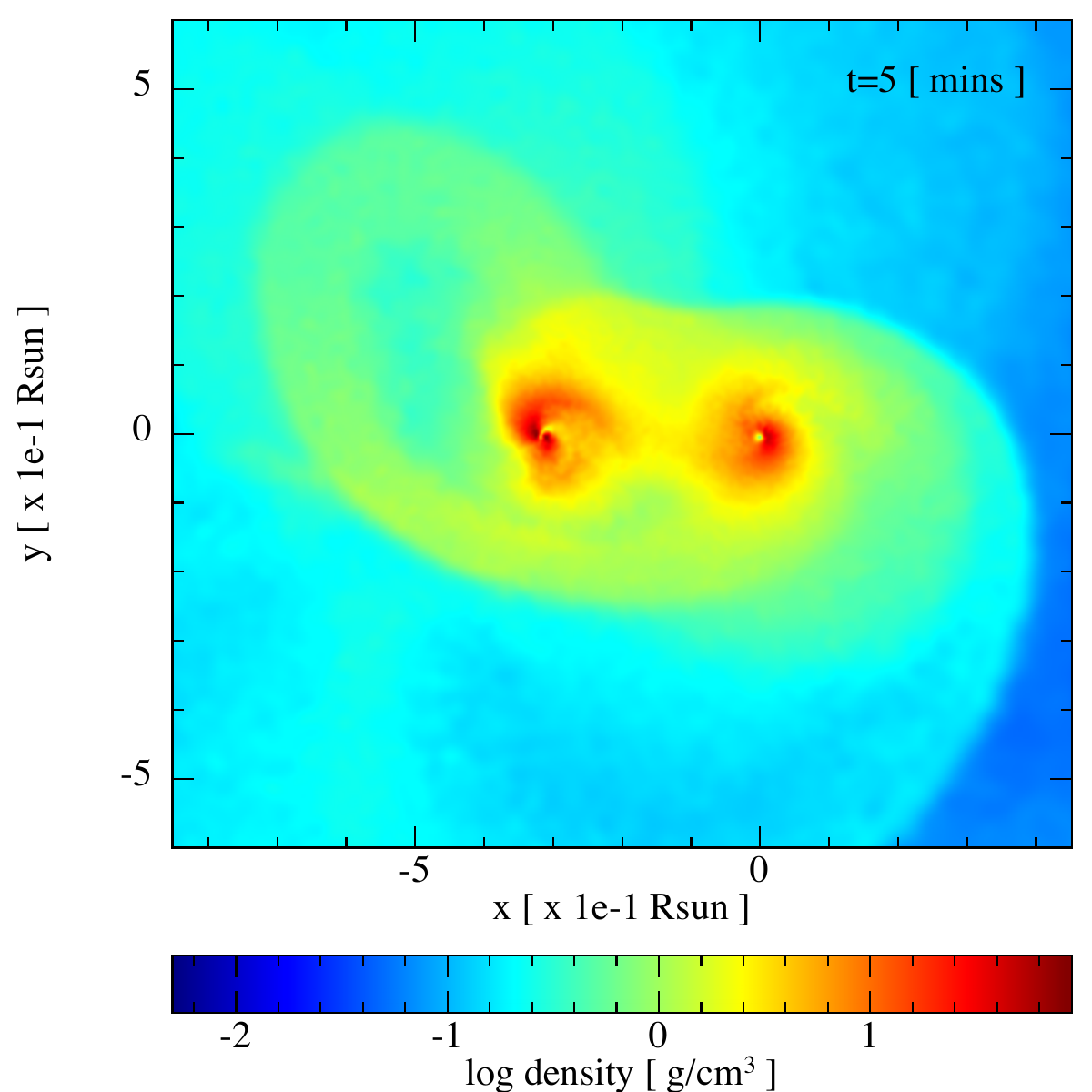}\includegraphics[width=0.33\hsize,clip]{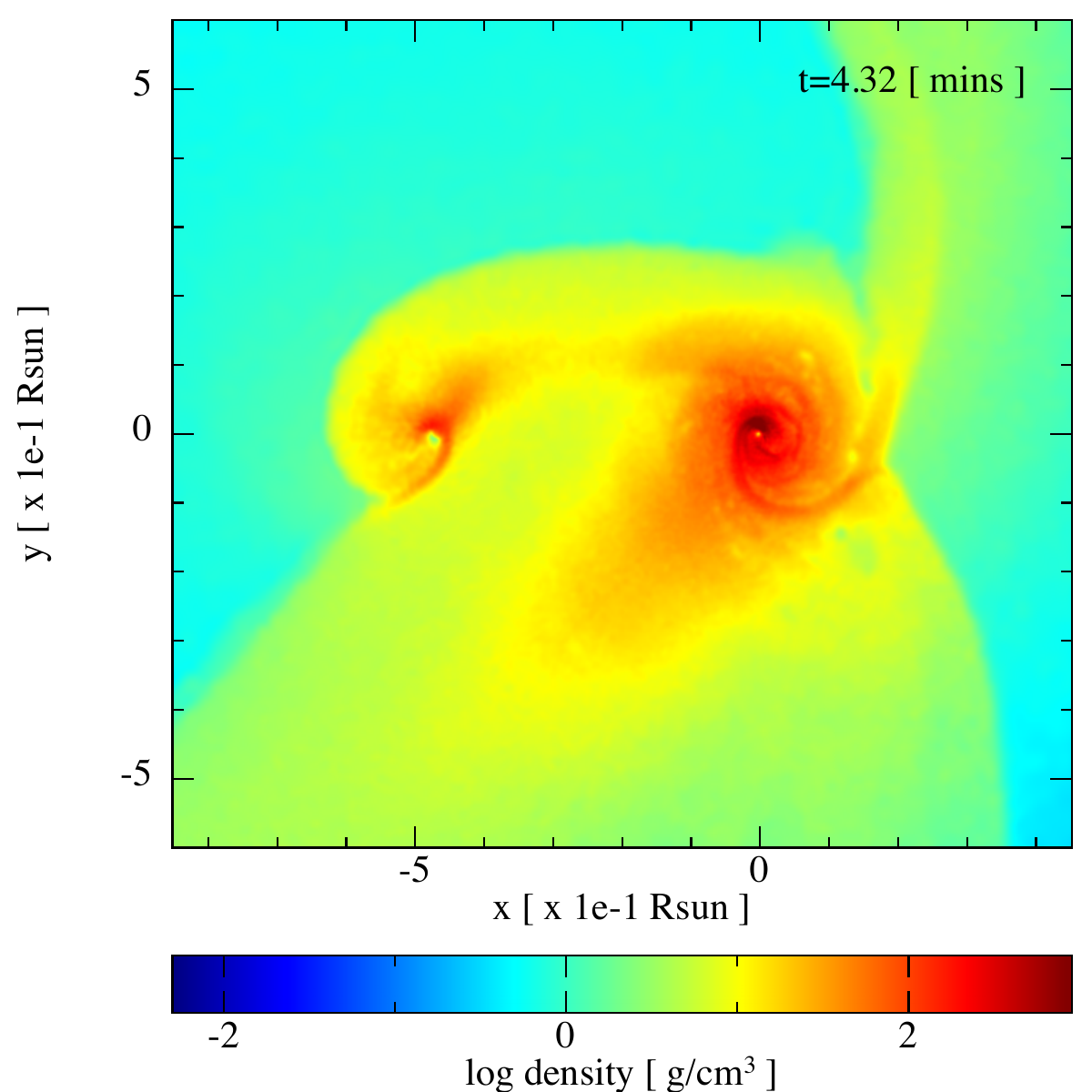}\includegraphics[width=0.33\hsize,clip]{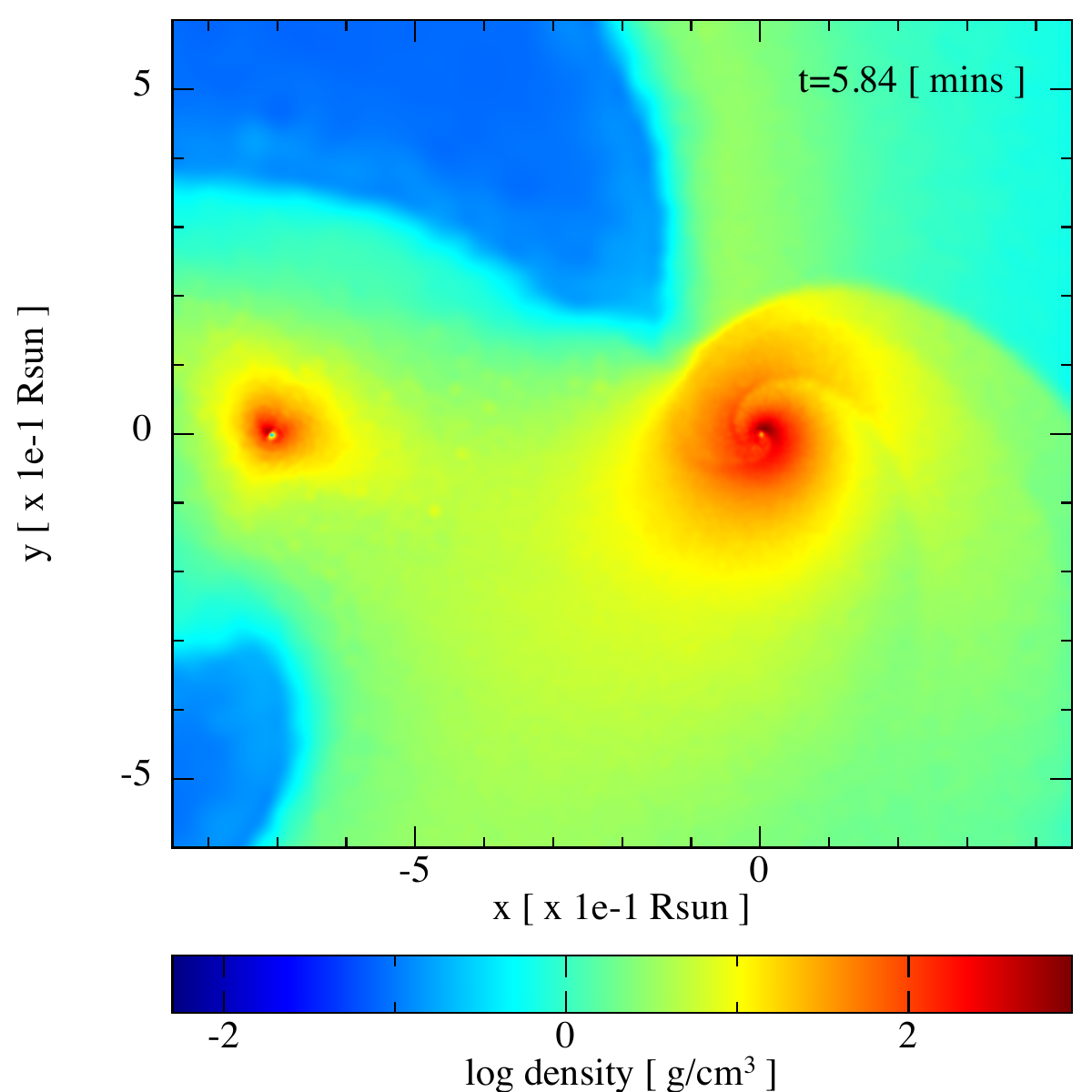}
    \caption{SPH simulations with a binary progenitor composed of a $1.4 M_\odot$ NS companion and a CO star, produced from different ZAMS stars of $15 M_\odot$ (left panel), $25 M_\odot$  (central panel), and $30 M_\odot$ (right panel). The orbital periods are $ 4.47$, $4.80$, and $4.88$ min, respectively. Each frame shows the mass density on the equatorial plane at selected times, with $t = 0$~s the instant of the SN shock breakout. The reference system is rotated to align the x-axis with the line joining the binary components and then translated to locate its origin at the position of the NS companion. The particles entering the NS capture region circularize around it, forming a thick disk already visible in all frames. Additionally, part of the SN ejecta is also attracted by the newborn NS accreting onto it and forming a disk structure. This figure was produced using the SNsplash visualization program \cite{2011ascl.soft03004P}.}
    \label{fig:3DSPH}
\end{figure*}

\section{Early evolution of the system}\label{sec:2}

\begin{figure*}
    \centering
   \includegraphics[width=0.49\hsize,clip]{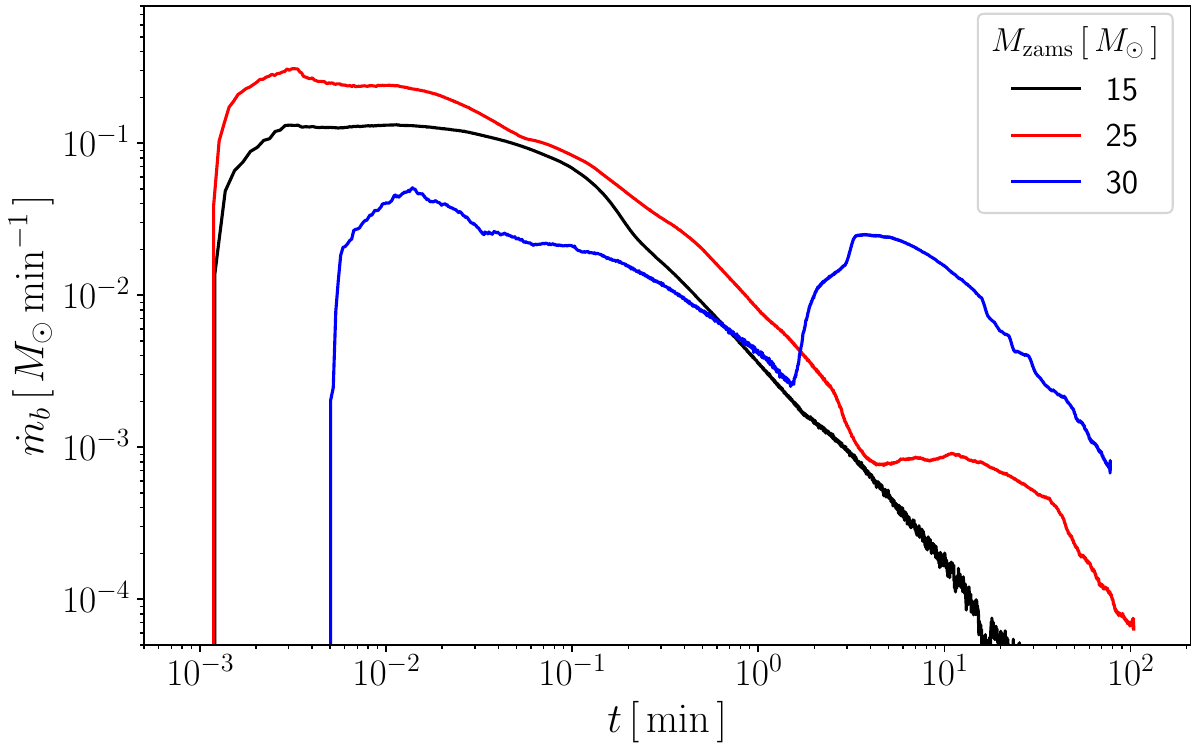}
   \includegraphics[width=0.49\hsize,clip]{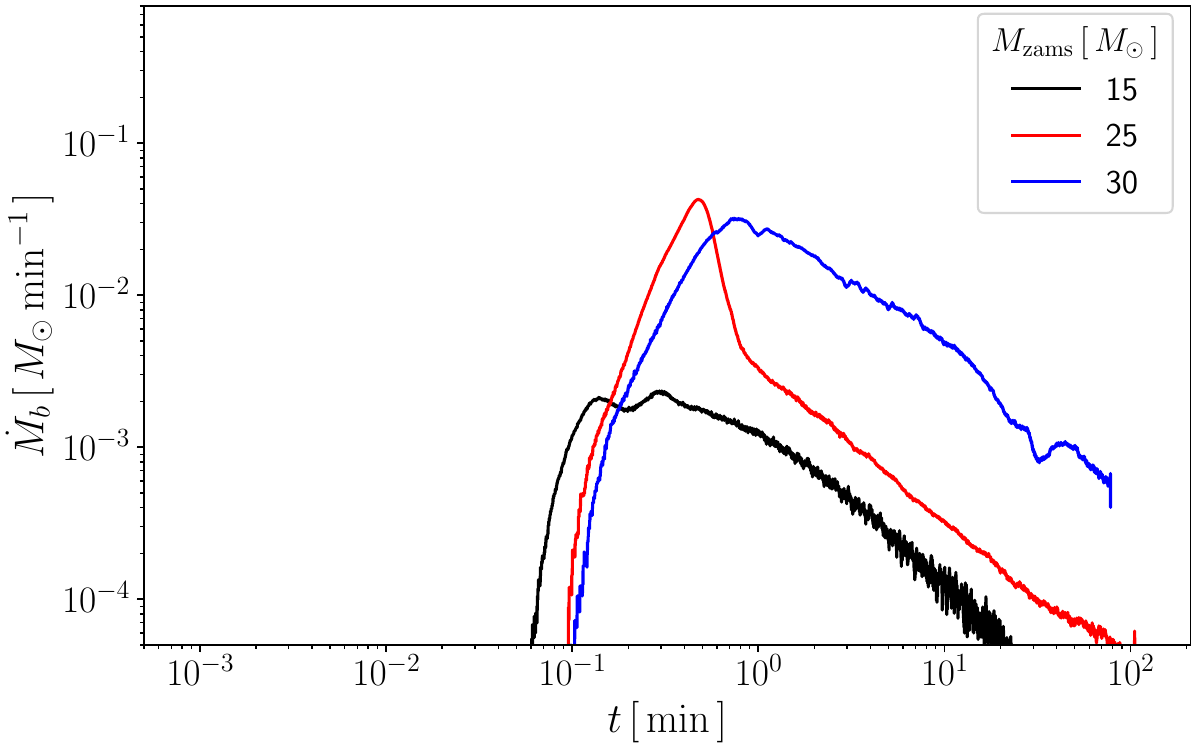}
    \caption{Baryonic mass accretion rate onto the newborn NS (left panel) and the NS companion (right panel) as a function of time, obtained from SPH simulations of the binary progenitor composed by a $1.4 M_\odot$ NS companion and a CO star, produced from different ZAMS stars: $15~M_\odot\, (E_{\rm sn}=1.08\times 10^{51}$~erg), $25 ~M_\odot\, (E_{\rm sn}=5.67\times 10^{50}$~erg), $30 ~M_\odot\, (E_{\rm sn}=7.85\times 10^{50}~$erg).}
    \label{fig:Mdots_Porb}
\end{figure*}

\begin{table}
    \centering
    \begin{adjustbox}{width=\textwidth}
    \begin{tabular}{c|cccccc}
    \hline
         $M_{\rm zams}\, (M_\odot)$& $M_{\rm CO}\,(M_\odot)$ &$m_b\,(M_\odot)$ & $m$ ($M_\odot$) &  $M_{\rm ej}\, (M_\odot)$ & $M\, (M_\odot)$& $P_{\rm orb}\, (\mathrm{min})$  \\
          \hline
             $30$ & $8.89$ & $1.75$ & $1.58$ & $7.14$ & $1.40$ & $4.88$ \\
         $25$ & $4.80$ & $1.80$ & $1.62$ & $ 3.00$ & $1.40$ & $4.80$\\
         $15$ & $3.00$ & $1.40$& $1.29$  & $1.60$ & $1.40$ & $4.47$ \\
          \hline\hline
    \end{tabular}
    \end{adjustbox}
    \caption{Properties of the CO-NS progenitor. The CO star is the pre-SN configuration, obtained from the evolution of a ZAMS star of mass $M_{\rm zams}$ using the KEPLER stellar evolution code \cite{2010ApJ...724..341H}. The pre-SN CO star mass is the sum of the mass of the outer layers expelled by the SN, $M_{\rm ej}$, and the collapsed iron core mass, assumed to be equal to the baryonic mass of the central remnant, the newborn NS, $m_b$. To estimate the newborn NS gravitational mass, $m$, we used the binding energy universal equation calculated in \cite{2015PhRvD..92b3007C}.}
    \label{tab:COstar}
\end{table}

We begin to obtain a realistic evolution of the accretion rate from numerical simulations of the accretion onto the newborn NS and its companion NS. For this task, we perform smoothed-particle-hydrodynamics (SPH) simulations with the SNSPH code \cite{2006ApJ...643..292F}, adapted to the occurrence of a SN explosion in a CO-NS binary as presented in \cite{2019ApJ...871...14B}, and extensively tested in previous studies of the BdHN scenario of long GRBs (see, e.g., \cite{2019ApJ...871...14B,2022PhRvD.106h3002B,2024ApJ...976...80B}, and references therein).

The SPH simulation begins when the SN shock front reaches the CO star surface; i.e., a 3D-SPH configuration is mapped to a 1D core-collapse SN simulation, as described in \cite{2018ApJ...856...63F}. By this time, the collapse of the CO star has formed a newborn NS of baryonic mass $m_b$ and gravitational mass $m$. The SN explosion ejects matter, and a part of it accretes both on the newborn NS and the NS companion. The latter has an initial baryonic mass $M_b$ and gravitational mass $M$. In the simulation, the newborn NS and the NS companion are modeled as point-like masses and interact only gravitationally with the SN particles and with each other. We allow the particles to increase their masses by accreting other particles from the SN material, following the algorithm described in \cite{2019ApJ...871...14B}.

Figure \ref{fig:3DSPH} shows snapshots of the mass density of the SN ejecta in the y-x plane (the binary equatorial plane) from three different sets of parameters. We performed simulations with an initial binary system formed by an $M = 1.4 M_\odot$ NS companion and a CO star that originated from selected ZAMS progenitors of $M_{\rm zams}=15$, $25$, and $30 M_\odot$. For each of these progenitors, Table~\ref{tab:COstar} summarizes some of the binary properties, including the baryonic mass $m_b$ of the newborn NS formed from the collapse of the iron core of the CO star, the mass ejected by the SN explosion, $M_{\rm ej}$, the gravitational mass $M$ of the NS companion, and the pre-SN orbital period. The latter is the minimum value for the pre-SN binary to have no Roche-lobe overflow.

The SN ejecta that the NS companion gravitationally captures forms a tail behind it, and circularizes, forming a thick disk. At the same time, the particles from the innermost layers of the SN ejecta that could not escape from the newborn NS gravitational field produce a fallback accretion onto the newborn NS. After a few minutes, the material in the disk around the NS companion is also attracted by the newborn NS, enhancing the accretion process onto the newborn NS. After about one orbital period, a disk has formed around the NS companion and the newborn NS, as shown in Figure~\ref{fig:3DSPH}.

Figure~\ref{fig:Mdots_Porb} shows the accretion rate onto the newborn NS and the NS companion obtained from SPH simulations for the three CO star progenitors used in this study. The accretion rate onto the newborn NS shows two prominent peaks. The second peak of the fallback accretion rate onto the newborn NS is caused by the influence of the NS companion \cite[see][for additional details]{2019ApJ...871...14B}. The accretion rate onto the NS companion shows a single-peak structure, accompanied by additional peaks of smaller intensity and shorter timescales. The small peaks are due to episodes of higher accretion as the NS companion travels into the ejecta and finds larger and lower-density regions as it continues to orbit.

\section{Neutron star evolution}
\label{sec:nsevolution}

To calculate the time evolution of the newborn NS and the NS companion structure during the accretion process, we have implemented a code that uses the {\sc RNS code} \cite{1995ApJ...444..306S}, with the quadrupole correction performed in \cite{2015PhRvD..92b3007C}. Given the EOS, the code calculates the stable, rigidly rotating NS configuration of equilibrium in axial symmetry, for a given baryonic mass and angular momentum, $J$.\footnote{In this work, we use both hadronic and SQS EOSs. The RNS code requires as input the pressure $p$ and energy density $e$. From them, it computes the log-enthalpy by integrating $d(\log h) = dp / (e + p)$, and then the particle density as $n_B = (e + p) \exp(-h) / m_p$, with $m_p = 939$ MeV. The integration is performed backward from the surface to the center. For hadronic EOSs, the integration starts at $\log h(p = 0) = 1$. For SQS EOS, the code was modified so that the integration starts at $\log h(p = 0) = e(p = 0) / (m_p n_B(p = 0))$.} The value of $M_b$ is updated using the baryonic mass accretion rate from the SPH numerical simulation (see Sec. \ref{sec:2}). The value of $J$ is obtained by angular momentum conservation \cite{2015ApJ...812..100B,2017PhRvD..96b4046C,2022PhRvD.106h3002B}
\begin{equation}\label{eq:Jdot}
\dot{J}= \tau_{\rm acc},
\end{equation}
where $\tau_{\rm acc}$ is the accreting  torque acting onto the stars. Our numerical simulations indicate that the infalling material, before being accreted, circularizes, forming a disk around the star. Therefore, the accreted matter exerts a (positive) torque onto the star. Denoting by $l$ the angular momentum per unit mass of the material at the inner disk radius, $R_{\rm in}$, the exerted torque onto the stars is given by 
\begin{equation}\label{eq:chi}
  \tau_{\rm acc} = \begin{cases}
      \chi\,l\,\dot{m}_b, & \text{torque onto the newborn NS,}\\
      \chi\,l\,\dot{M}_b, & \text{torque onto the NS companion},
  \end{cases}
\end{equation}
where $\chi \leq 1$ is an efficiency parameter of angular momentum transfer. The specific angular momentum $l$ is
\begin{equation}
    l = \begin{cases}
    l_{\rm isco} ,& \text{if } R_{\rm in}\geq R, \\
    \Omega R^2,              & \text{if } R_{\rm in}< R,
\end{cases}
\end{equation}
where $l_{\rm isco}$ is the value of the innermost stable circular orbit, while $R$ and $\Omega$ are the radius and angular velocity of the NS, which are obtained from the numerical solution of the Einstein equations directly from the {\sc RNS code}. It is worth noting that for this work, we have neglected the torque due to the magnetic braking mechanism \cite[see previous works, e.g.,][]{2022PhRvD.106h3002B}, as we are interested in the early evolution of the star when the accretion rate is high and the magnetospheric radius is within the star \cite{Becerra:2024ond}.
%
%
%
%
%
%


\section{Equations of state and conversion to quark matter}
\label{sec:eos}

The EOS of hadronic matter is described within the relativistic mean field model SFHo proposed in \cite{Steiner:2012rk}, and which has been extended to include hyperons and delta resonances \cite{Drago:2014oja,Drago:2015cea}. Here, we adopt the same parametrization used in \cite{Burgio:2018yix,DePietri:2019khb,Bombaci:2020vgw}: the scalar $\sigma$ meson-hyperon coupling constants have been fitted to reproduce the potential depth of the corresponding hyperon at the saturation nuclear matter density, while the SU(6) symmetry relations are assumed for the coupling between hyperons and vector mesons. The couplings between mesons and  delta resonances are $x_{\sigma \Delta}=1.15$ and $x_{\omega \Delta}=x_{\rho \Delta}=1$ (\cite{Drago:2015cea,DePietri:2019khb,Drago:2014oja,Burgio:2018yix}, for a detailed discussion).

While the NS is, at least initially, cold, the newborn NS is born as a hot proto-NS (PNS). Qualitatively, the thermal evolution of a PNS can be approximated as follows \cite{1997PhR...280....1P,Roberts_2012,Camelio_2017}: after the SN explosion, the PNS is characterized by an entropy per baryon of $S/A\sim 1$--$1.5$, and neutrinos are trapped, resulting in a lepton fraction of $Y_{Le}=0.3$--$0.4$. After $\sim 10$ s, the PNS core reaches its maximum entropy per baryon $S/A\sim 1.5$--$2$, while the lepton fraction reduces to $Y_{Le}\sim 0.2$--$0.3$ due to neutrino diffusion.
During the next $\sim (30$--$60)$ s, the PNS rapidly cools by emitting neutrinos, and the core temperature drops to a few MeV, eventually reaching the cold, neutrinoless $\beta$-equilibrium NS configuration. Since the simulation does not evaluate the thermal and lepton number evolution of the system, assumptions have to be made. We are interested in the conversion of the two NSs into SQSs; therefore, a suitable approach would be to establish the conditions that NSs typically exhibit when the conversion begins. However, in principle, depending on the progenitor star mass, the explosion energy, and the initial angular momentum of the newborn NS, the deconfinement conditions could be either reached after some minutes, when also the newborn NS is cold, or after a couple of seconds, when it is still hot.

Thus, we will use two different conditions for the hadronic matter: i) $T=0$, neutrinoless $\beta$-equilibrium hadronic EOS (EOS1); and ii)  $S/A=1$, neutrinoless $\beta$-equilibrium EOS (EOS2) to investigate the systematics due to the finite temperature. Note that the mass accretion does not significantly heat the cores of both NSs as a result of efficient atmosphere cooling due to neutrino emission (see \cite{Becerra:2016gdc,Becerra:2017olp}, for details). Thus, even during the mass accretion, the compact stars can be safely described with a cold EOS.

\begin{figure}
    \centering
   \includegraphics[width=\hsize,clip]{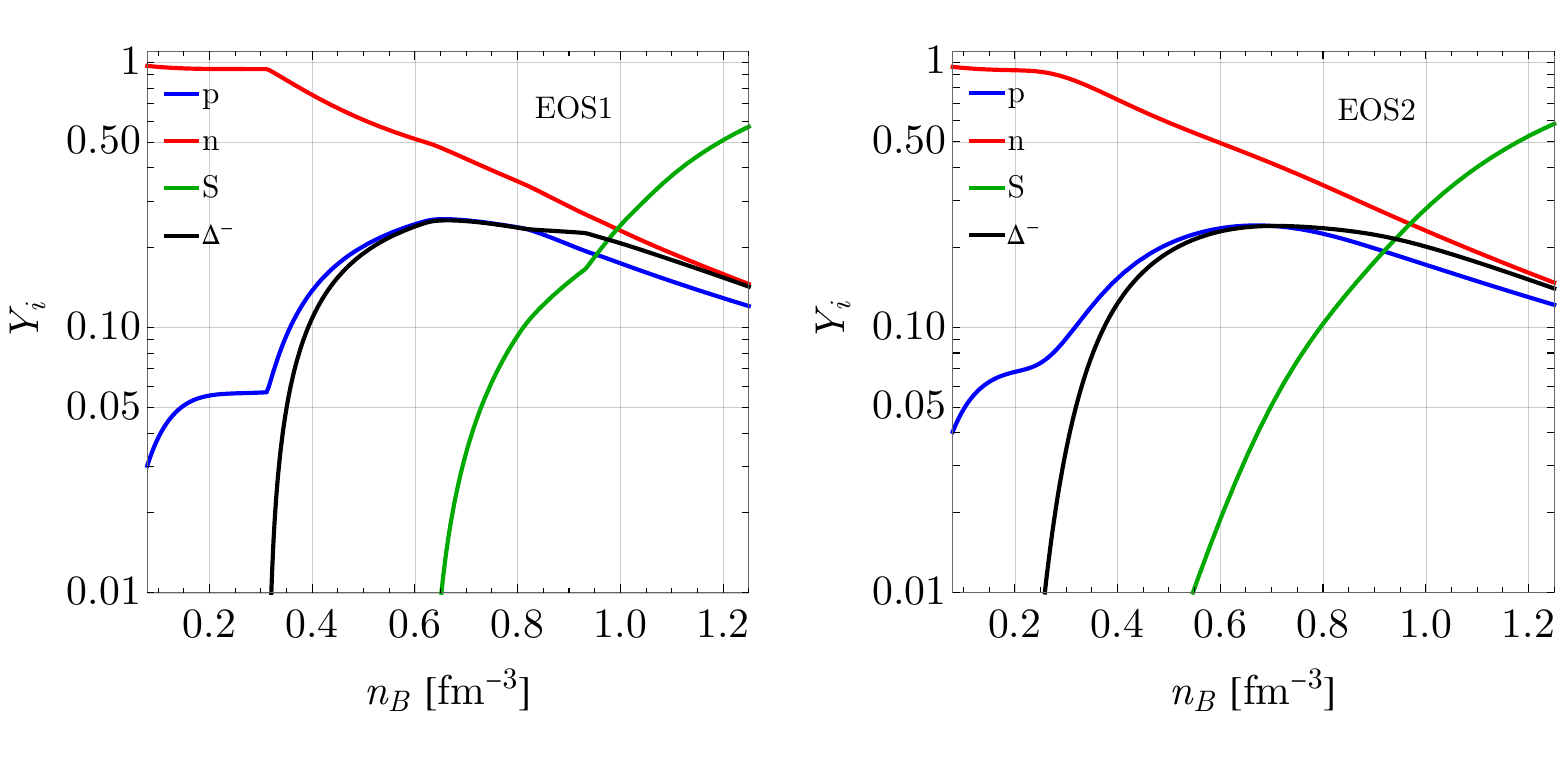}
    \caption{Fractions of neutrons, protons, $\Delta^-$ and strangeness ($Y_S$) as functions of the baryon density for the two EOSs here adopted.}
    \label{fig:fractionsEOS1} 
\end{figure}

For the following discussion, we display in Fig.~\ref{fig:fractionsEOS1} the particle fractions $Y_i=n_i/n_B$ and the total strangeness fraction $Y_S$ as functions of the baryon density for the two EOSs described above. 
Since we assume the absolute stability of $\beta$-equilibrium SQM in bulk (Bodmer-Witten hypothesis), ordinary $\beta$-equilibrium hadronic matter is always metastable. Assuming a first-order transition, a system in a metastable phase (hadronic phase) converts to the stable phase (SQM) after the nucleation process, i.e., the formation of the first seed of SQM. The deconfinement is mediated by the strong interaction, whose typical timescale is much smaller than that of the weak interaction. As a result, during the formation of the SQM droplet, the weak interaction does not have time to change the flavor composition of matter, and the flavors are therefore conserved. The SQM, therefore, begins to appear when thermodynamic conditions are such that there is a certain amount of strangeness in the hadronic system. 
A comprehensive discussion of the thermodynamic conditions under which quark nucleation occurs in both cold and hot stars can be found in \cite{Mintz:2009ay,Bombaci:2016xuj,Guerrini:2024gzu,Guerrini:2025mxx}.  
For simplicity, in this work, we will use the approximate approach presented in \cite{DePietri:2019khb,Bombaci:2020vgw}, where SQM is assumed to nucleate when the average distance between strange quarks in the hadronic phase is of the order of the average distance of nucleons in nuclear matter, namely the strangeness density is of the order of nuclear matter saturation density. The reason is that strange quarks should be close enough to mutually interact to help the nucleation of deconfined quark matter. As shown in \cite{DePietri:2019khb}, such a condition is satisfied when the total strangeness fraction is $Y_S\sim 0.2$--$0.3$. Thus, we will assume that when the critical density $n_{\rm crit}$ at which $Y_S\sim 0.2$--$0.3$ has been reached during the temporal evolution of the two compact objects, the first seed of SQM is formed and the conversion to SQS starts. Note that in this work $Y_S\equiv n_s/n_B$ is equal to the number density of quark $s$ divided by the baryon number density (e.g., $\Lambda$ has $S=1$).


The conversion process is rather complex with an initial fast stage, lasting a few milliseconds, in which almost the whole star converts due to the development of hydrodynamical instabilities followed by a slow diffusive regime, which can last tens of seconds if the hadronic material is cold, but can be much faster if it is already hot, due to convective instabilities. During that time, the remaining external layers of the hadronic star continue to convert to SQM \cite{Drago:2015fpa,Drago:2015dea}.  Since we typically observe the accretion of a compact object over minutes, we assume that once the critical density is reached, the NS instantaneously converts to an SQS. After that instant, the matter composing the star must be described by an SQM EOS.  
We stress again that this criterion is rather qualitative, and we leave a more detailed analysis for a future study.  

We explore the effect of the SQM EOS by using two different models. The first is a MIT bag model with perturbative corrections (quark EOS1) \cite{Weissenborn:2011qu}, with the parameters $B^{1/4}=137.5$ MeV and $a_4=0.75$ (corresponding to $\alpha_s=\pi/2 \times 0.25$). The second one (quark EOS2) is a CFL bag model \cite{Alford:2004pf,Alford:2007xm,Weissenborn:2011qu}, with $B^{1/4}=135$ MeV, $a_4=0.7$ and $\Delta=80$ MeV \cite{Bombaci:2020vgw}. In both cases, $m_u=m_d=0$ and $m_s=100$ MeV. In principle, the system is hot during the conversion, and the SQS becomes cold after some tens of seconds \cite{Drago:2015fpa,Drago:2015dea}. Given that the typical timescale of accretion is of the order of minutes, we assume the cold EOS approximation for the SQS after its formation.  

Table~\ref{tab:EoS_models} shows the main properties of the EOSs used in this paper. It summarizes the maximum gravitational mass for both the static, $M_{\rm max}^{J=0}$, and Keplerian, $M_\mathrm{max}^K$, configurations, and the minimum mass required for the static star to reach the critical particle density, $M(n_{\rm crit}^{J=0})$.

\begin{table}
    \centering
    \begin{tabular}{c|ccc}
    \hline
       EOS  & $M^{J=0}_{\rm max}$ & $M^K_{\rm max}$ & $M^{J=0}(n_{\rm crit})$  \\
        & ($M_\odot$) & ($M_\odot$) & ($M_\odot$)  \\ \hline \hline
        EOS1: SFHo-H$\Delta$  & $1.585$ & $1.884$& $1.545$--$1.566$ \\
         ($T=0$)  &  &  & \\ \hline 
        EOS2: SFHo-H$\Delta$ &  $1.576$  &  $1.850$ & $1.519$--$1.550$\\
          ($S/A=1) $  &  &  &  \\ \hline
        Quark EOS1& $2.116$ & $3.042$ & --\\
     unpaired Bag model    &  &  &  \\ \hline
        Quark EOS2   & $2.586$ & $3.719$ &-- \\
        CFL Bag model   &  &  &  \\
            \hline \hline
    \end{tabular}
    \caption{Maximum gravitational mass for the static and Keplerian (mass-shedding) sequences, $M_\mathrm{max}^{J=0}$ and $M_\mathrm{max}^K$, and the minimum mass required for the static star to reach the critical particle density, $M(n_{\rm crit}^{J=0})$, for the EOS models used.}
    \label{tab:EoS_models}
\end{table}

\section{Results}
\label{sec:results}

\begin{figure*}
    \centering
   \includegraphics[width=\hsize,clip]{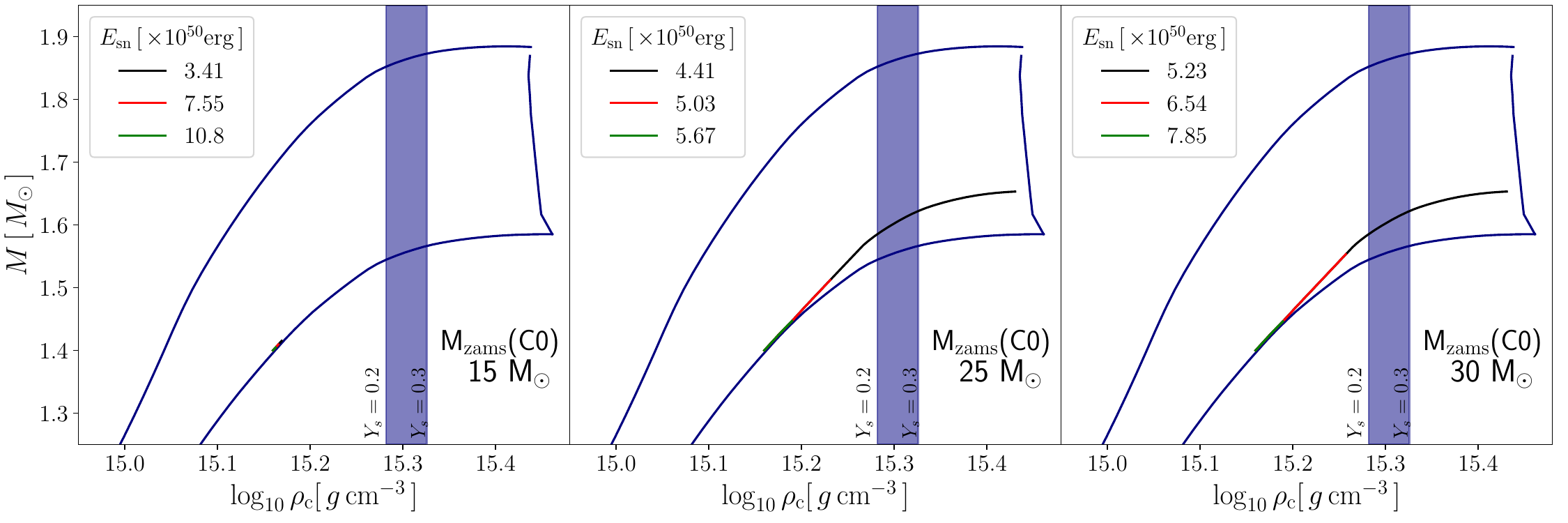}
    \caption{Evolution in the central density-gravitational mass plane for an initially non-rotating NS companion with $1.4~M_\odot$ using hadronic EOS1 (cold EOS). The CO star originates from a ZAMS progenitor with $M_{\rm zams}=$15 (left panel), $25$ (central panel), and $30~M_\odot$ (right panel), with an initial binary period of $4.47$~min, $4.8$~min, and $4.88$~min, respectively. Different colors correspond to different SN energies, $E_{\rm sn}$. The dark-blue solid lines bound the NS stability region. The colored band highlights the region where the strangeness fraction reaches the critical regime of $0.2$--$0.3$.}
    \label{fig:Mrho_NS}
\end{figure*}

\begin{figure*}
    \centering
   \includegraphics[width=\hsize,clip]{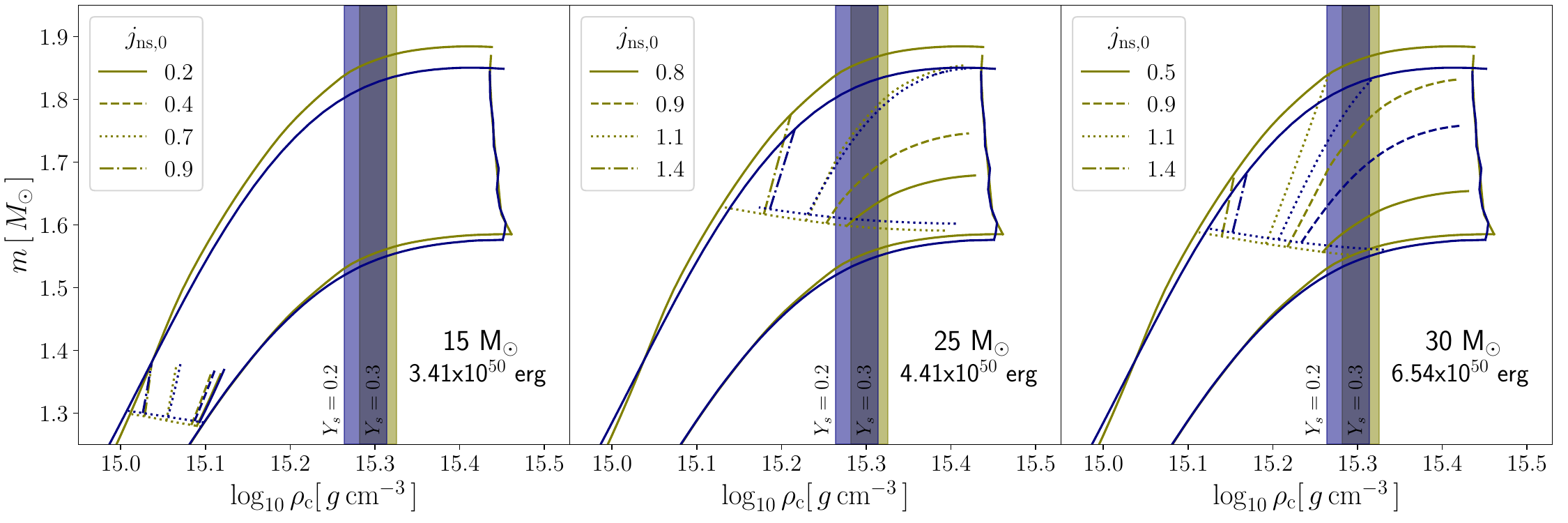}
    \caption{Same as Fig.~\ref{fig:Mrho_NS} for the newborn NS with different initial angular momentum along a sequence of constant baryonic mass: $m_b= 1.4$ (left), $1.85$ (center), and $1.75~M_\odot$ (right) for the CO star from progenitors with $M_{\rm zams}= 15$ (left panel), $25$ (central panel) and $30~M_\odot$ (right panel, respectively. The SN energy $E_{\rm sn}$ is fixed for the different CO star progenitors. Yellow color corresponds to the hadronic EOS1, and dark blue color to the hadronic EOS2.}
    \label{fig:Mrho_nuNS}
\end{figure*}

We start by analyzing the evolution of a $1.4~M_\odot$ NS companion for the three CO progenitors with $M_{\rm zams}=15$, $25$, and  $30~M_\odot$. For each progenitor, we fix three possible values of the SN explosion energy $E_{\rm sn}$ and set the initial rotational frequency of the compact star to zero. The chosen SN explosion energies are such to be sufficient to unbind the outer layers of the CO star, and produce ejecta velocities from $1,000$ km s$^{-1}$ of the innermost layers to a few $10,000$ km s$^{-1}$ of the outermost layers (see, \cite{2019ApJ...871...14B}, for details). We end the simulation when the bound mass to the system gets lower than $10^{-3} M_\odot$, or when any of the NSs reaches the Keplerian (mass-shedding) or secular instability. When one of the NSs reaches the critical density for SQM deconfinement, we continue the simulation with the new interior physics constitution to explore whether the other NS can also reach the conversion condition. It is worth noting that this assumption neglects the effects on the system of energy release resulting from the conversion.

In Fig.~\ref{fig:Mrho_NS}, we display the evolution of the NS gravitational mass and central density due to the accretion using the cold EOS (EOS1): as matter falls onto the compact star, its central density and rotational frequency increase (see green, red, and black lines). One can notice that the smaller the $E_{\rm sn}$, the larger the amount of baryonic mass accreted and the larger the increment of the central density. 

In the same plots, we also display the non-rotating, secular instability, and Keplerian sequences, along with the density range where the strangeness fraction reaches values of $0.2$--$0.3$ for the adopted EOSs. For the CO progenitor with $M_{\rm zams}=15~M_\odot$, the NS companion does not reach conditions for nucleation of SQM for any of the $E_{\rm sn}$ values adopted. The mass accretion rate is too low for this progenitor. Instead, for the other two massive CO progenitors, $M_{\rm zams}=25$  and 30~$M_\odot$, and for the smallest value of $E_{\rm sn}$ considered in each case, the NS star core reaches particle densities above $n_{\rm crit}$. In our scenario, those NSs would then rapidly convert to SQSs. 

\begin{figure*}
    \centering
   \includegraphics[width=\hsize,clip]{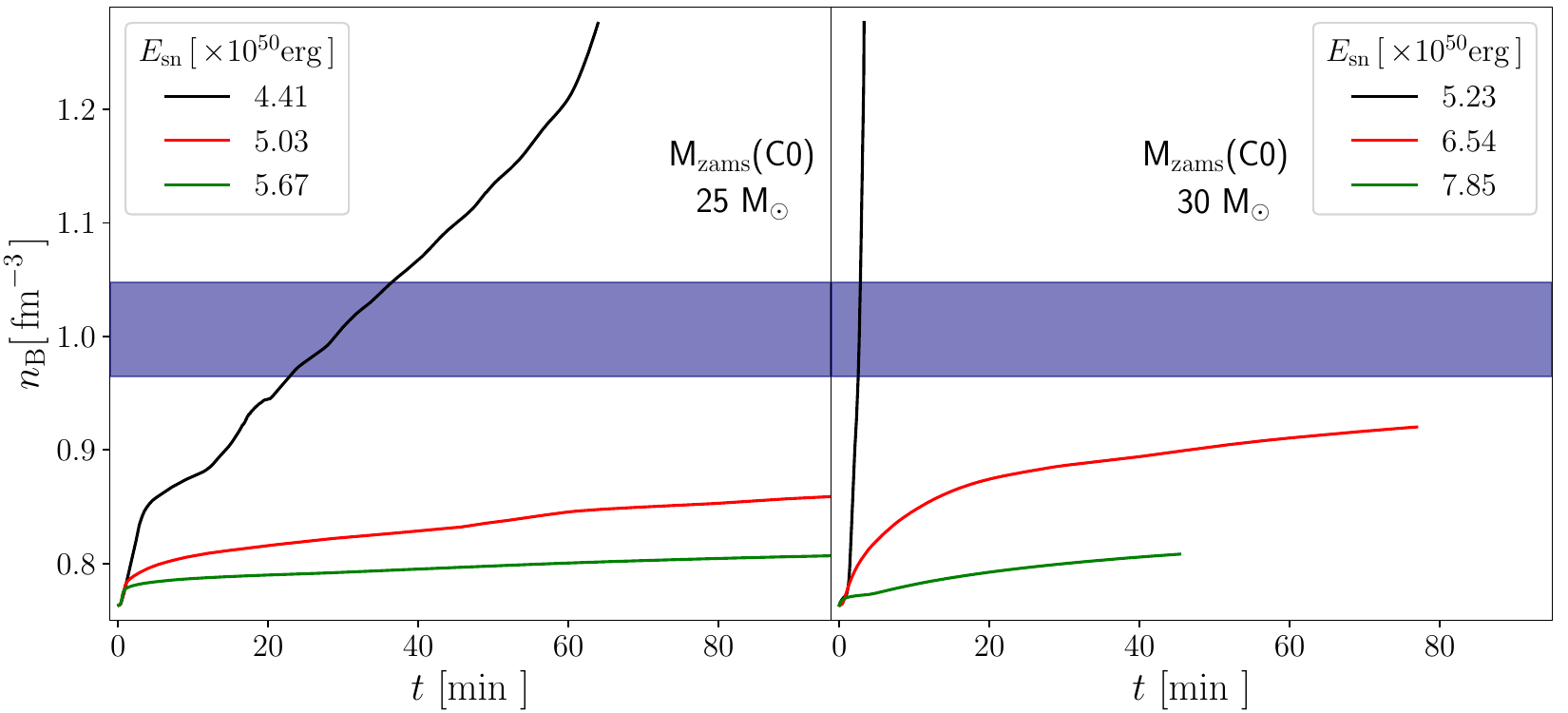}
   \includegraphics[width=\hsize,clip]{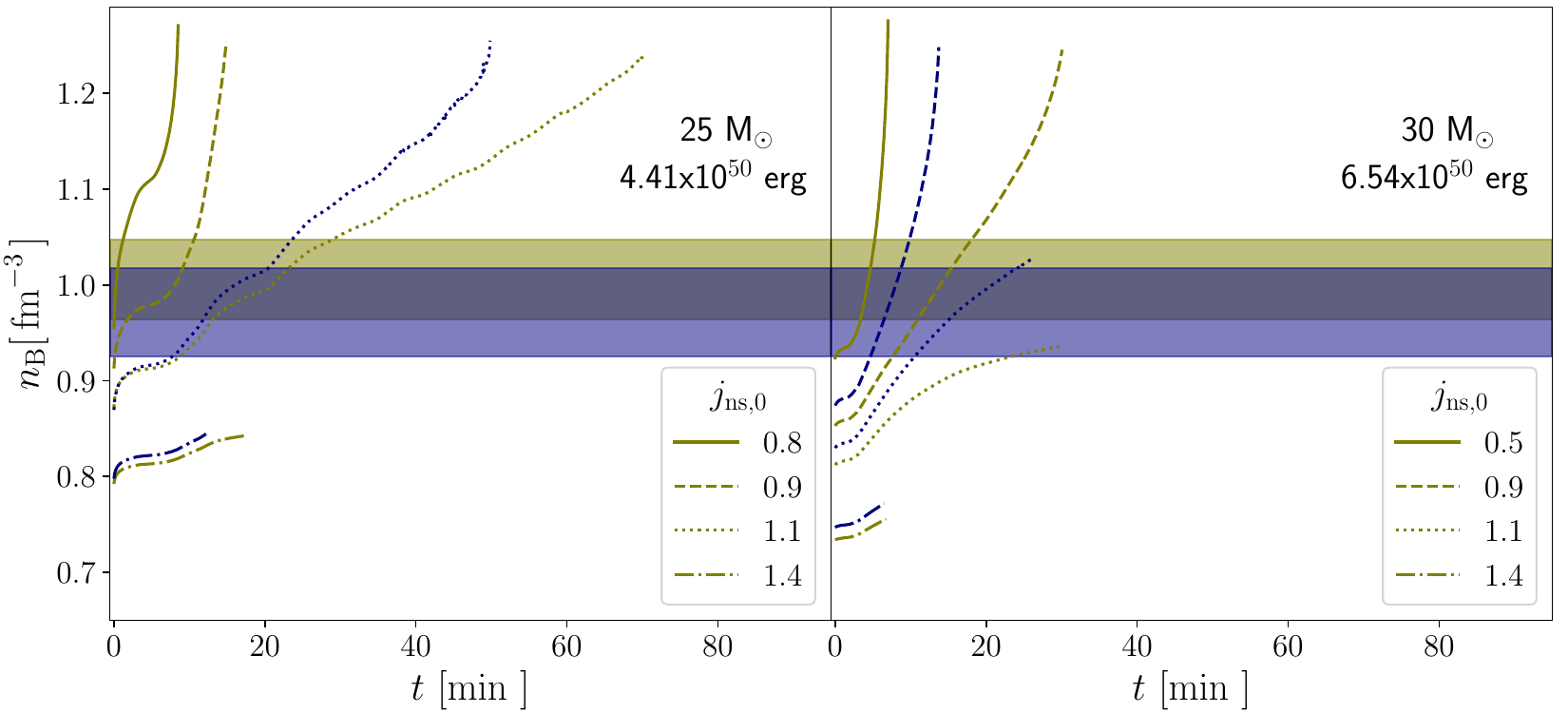}
    \caption{Upper panel: Time evolution of the central baryon particle density of the NS companion, for the simulations shown in Fig.~\ref{fig:Mrho_NS}; i.e., fixed initial angular momentum (zero) and varying SN kinetic energy. Lower panel: Time evolution of the central baryon particle density of the newborn NS, for the simulations shown in Fig.~\ref{fig:Mrho_nuNS}; i.e., fixed SN kinetic energy and varying initial stellar angular momentum. We have excluded the baryon density in the case of the CO star from the $M_{\rm zams} = 15 M_\odot$ because in that case none of the NSs reached the critical density for SQM deconfinement (see Figs.~\ref{fig:Mrho_NS} and \ref{fig:Mrho_nuNS}).}
    \label{fig:nBTime_NS}
\end{figure*}

In Figs.~\ref{fig:Mrho_nuNS} and \ref{fig:nBTime_NS} (lower panel), for the newborn NS, we have fixed $E_{\rm sn}$ to ($3.41$, $4.41$ and $5.23)\times 10^{50}$~erg, for the systems with CO star with $M_{\rm zams} = (15$, $25$ and $30$) $M_\odot$, respectively. We have considered different possible values of the initial specific angular momentum, $j_{\rm ns,0}\equiv c J_{0}/(G M_\odot^2)$, along a sequence of constant baryonic mass according to the remnant object formed by the collapse of each CO star (see Table~\ref{tab:COstar}). In the configurations with initial fast-rotating conditions (e.g., with $j_{{\rm ns},0}>1.2$, which corresponds to a rotation frequency of $\approx 1$ kHz), the accretion is more effective in speeding up the star rather than in compressing it, so the central density increases slowly while the mass and angular momentum increase rapidly. Under these conditions, the newborn NS reaches the Keplerian (mass-shedding) instability limit rather than the critical point for quark deconfinement. In the case of a slower rotating initial configuration, the central density increases more significantly, allowing the NS to convert to a SQS (see dashed and dotted lines of Fig.~\ref{fig:Mrho_nuNS}), as in the configuration with the CO star from the $M_{\rm zams} = 25 M_\odot$ and $30~M_\odot$ progenitor. It is worth noticing that when the initial angular momentum of the star is too small, the remnant object is born with a central particle density that exceeds the critical value for quark deconfinement. In such cases, our approach cannot be applied, as any conclusions on deconfinement should be properly investigated before the SN explosion or during the early evolution of the hot PNS.

Figure~\ref{fig:nBTime_NS} shows the time evolution of the central baryonic particle density for the simulations of Figs.~\ref{fig:Mrho_NS} and \ref{fig:Mrho_nuNS}, in the cases where any of the NSs reach conditions to convert into an SQS. The figure clearly shows that, for a fixed SN kinetic energy, the conversion of the newborn NS into a SQS is favored over that of the NS companion.

\begin{figure*}
    \centering
    \includegraphics[width=0.49\hsize,clip]{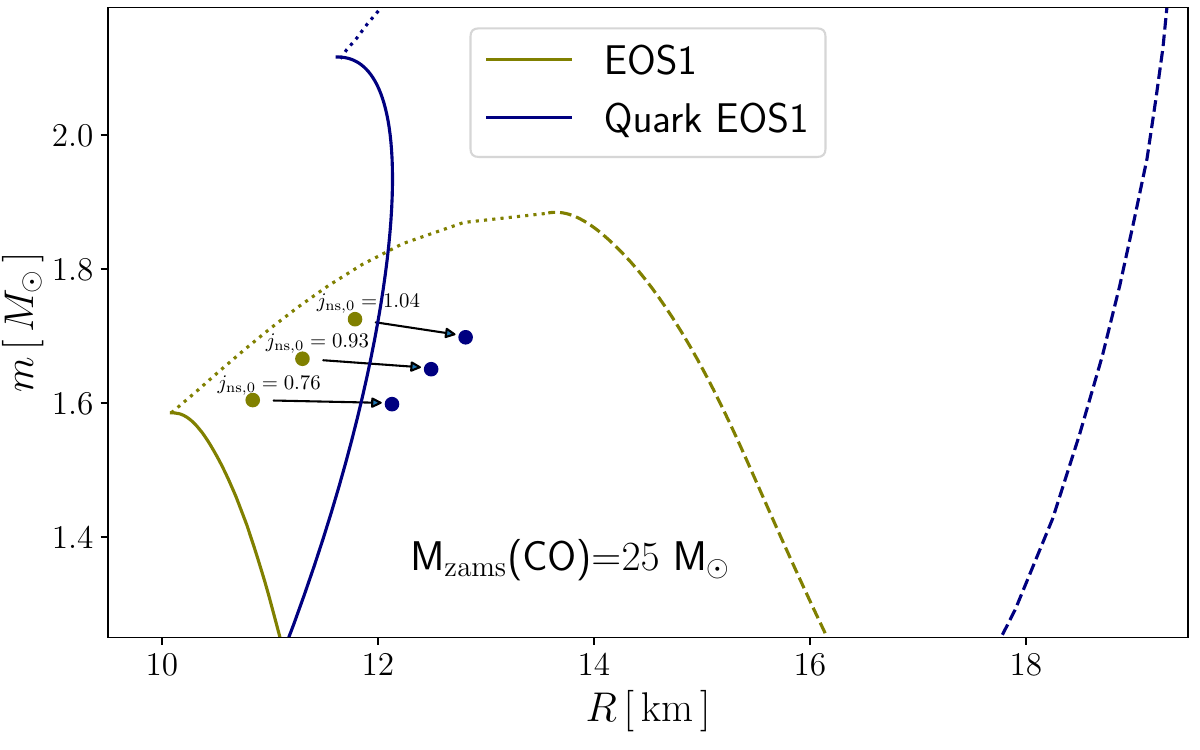}
    \includegraphics[width=0.49\hsize,clip]{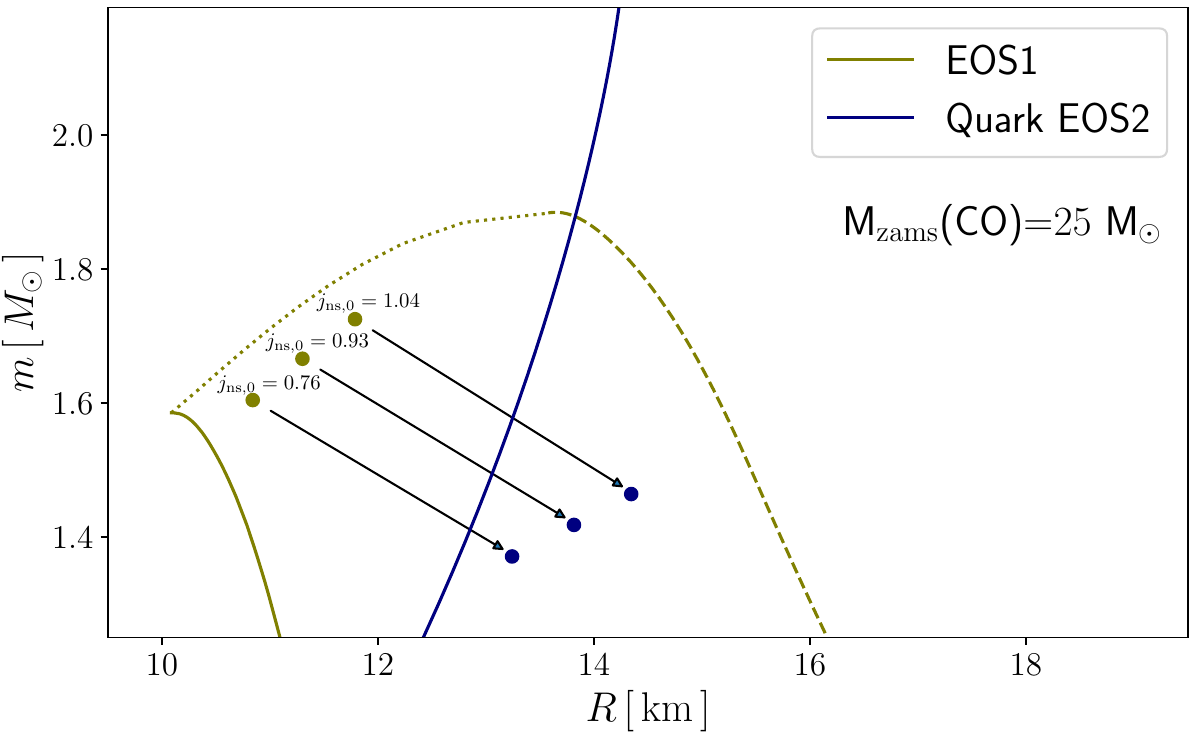}
    \includegraphics[width=0.49\hsize,clip]{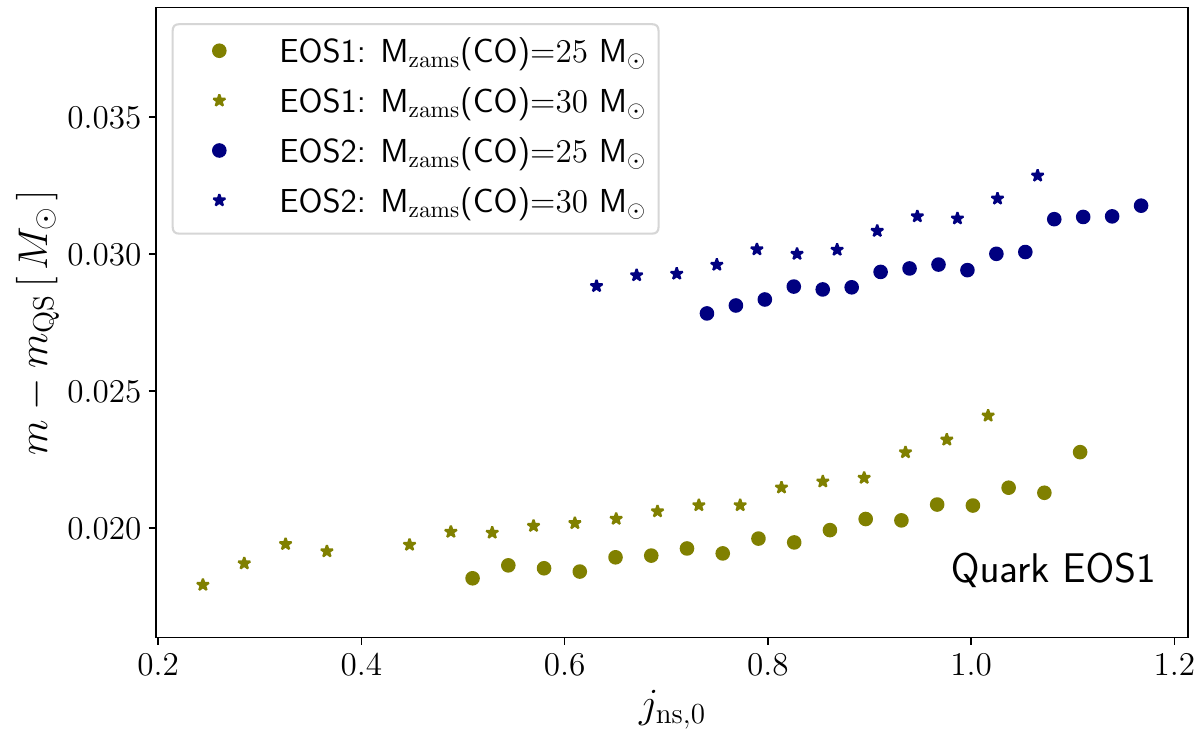}
    \includegraphics[width=0.49\hsize, clip]{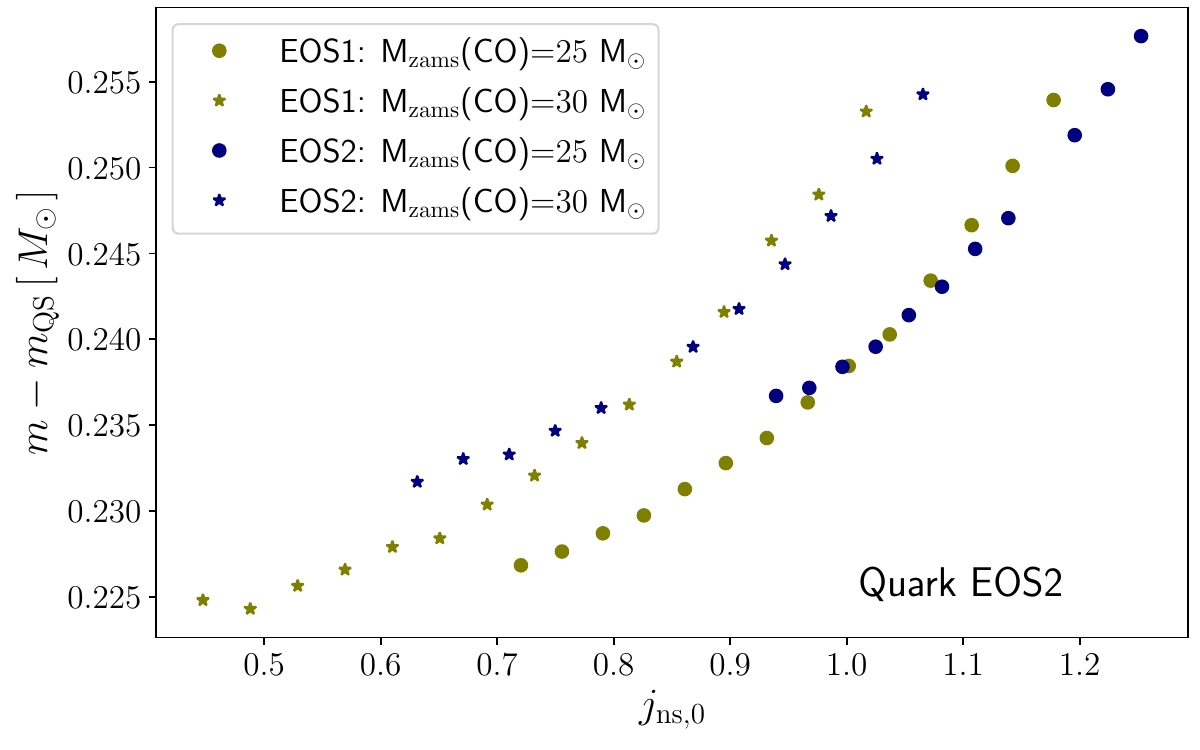}
    \caption{Upper panel: Gravitational mass and equatorial radius plane for the EOS1 model and quark EOS1 (left) and quark EOS2 (right), modeling NS and SQS star, respectively. The stability region is bounded in colored solid lines corresponding to the non-rotating (solid line), secular instability (dotted line), and Keplerian (dashed line) sequences. The dark-blue points represent the mass of the SQS with the same baryon number and angular momentum of the newborn NS when it reaches the critical density for quark deconfinement during accretion (yellow points). Generally, it reduces its gravitational mass and increases its radius. The  CO star originates from a progenitor with $M_{\rm zams}=$25~$M_\odot$ and the kinetic energy of the SN explosion is $\sim 3.4\times 10^{50}$~erg. Lower panel: Change of gravitational mass when the newborn NS converts into a SQS, as a function of the initial stellar angular momentum, for the quark EOS1 (left) and the quark EOS2 (right). The energy of the SN explosion corresponds to the values shown in Figure~\ref{fig:Mrho_nuNS}. The range of the initial angular momentum corresponds to rotating frequencies between $500$ and $1100$ Hz.}
    \label{fig:Qs}
\end{figure*}

In the present scenario, when any NS reaches central particle densities above $n_{\rm crit}$, it rapidly converts to SQSs. The upper panel of Fig.~\ref{fig:Qs} depicts the conversion in the gravitational mass and equatorial radius plane in some selected cases. We calculated the equivalent configuration using the quark EOS1 (left plot) and the quark EOS2 (right plot), maintaining the same angular momentum and baryonic mass when the star reaches a central strangeness fraction $Y_S=0.2$. The resulting SQS has a smaller gravitational mass and a larger equatorial radius, in agreement with the analysis of \cite{Drago:2020gqn}. The lower panel of Fig.~\ref{fig:Qs} shows the change in the gravitational mass of the newborn NS as a function of its initial angular momentum, in the case of the two more massive CO progenitors investigated ($M_{\rm zams}= 25$ and $30~M_\odot$). Assuming the quark EOS1, the change in the gravitational mass is about $0.02$--$0.025~M_\odot$ for the cold EOS and approximately $50~\%$ greater for the hot EOS. For the quark EOS2 case, the change in the gravitational mass is about $0.22$--$0.25~M_\odot$.
These values correspond to a conversion energy of $\approx 4 \times10^{52}$ erg and $\approx 4 \times10^{53}$ erg for the quark EOS1 and quark EOS2, respectively. The larger conversion energy associated with the quark EOS2 is due to its higher binding energy relative to that of the quark EOS1. The energy released in the conversion can lead to the emission of highly energetic electromagnetic transients (see e.g. \cite{Drago:2015dea} for a discussion on short and long gamma-ray-bursts) which we will discuss in a forthcoming paper.

Table~\ref{tab:outcome} summarizes the final fate of the simulations performed. In particular, for a given set of system parameters (mass of the CO star, the newborn NS, and the NS companion, orbital period, and SN explosion energy), we show the fate of the newborn NS and of the NS companion by the end of the simulation, hence, the remnant binary. 
It is interesting to note that, depending on the initial conditions, the astrophysical systems we are investigating can produce NS-NS, NS-SQS, and SQS-SQS systems. If those binaries remain bound, they could eventually produce compact star merger events whose gravitational and electromagnetic signals strongly depend on the composition of the two stars. In particular, in the two-families scenario, the progenitor of GW170817 has been interpreted as being 
a NS-SQS system \cite{Burgio:2018yix}. The results of Table~\ref{tab:outcome} 
show possible astrophysical situations in which such a mixed system could have been formed.

\begin{table}
    \centering
\begin{adjustbox}{width=\textwidth}
    \begin{tabular}{p{2.2cm}|c|ccc|ccc} 
    \hline 
    \multicolumn{8}{c}{\textbf{Initial system}} \\ \hline
      $M_{\rm CO}$ ($M_\odot$)  & $3$ ($M_{\rm zams} = 15~M_\odot)$  & \multicolumn{3}{c|}{$4.8$ ($M_{\rm zams} = 25~M_\odot)$} & \multicolumn{3}{c}{ $8.9$ ($M_{\rm zams} = 30~M_\odot)$}  \\
       $m_b$ ($M_\odot$)   & $1.40$  &  & $1.80$ & & & $1.75$&  \\ 
       $M$ ($M_\odot$)   & $1.40$  &  & $1.40$ & & & $1.40$ &  \\ 
       $P_{\rm orb}$ (min) & $4.47$  &  & $4.80$ & & & $4.88$&  \\ 
       \hline 
      \multicolumn{8}{c}{\textbf{Components fate}} \\
      \hline
       $E_{\rm sn}$ ($10^{50}$~erg)  & $3.41$--$10.8$ &$4.41$&$5.03$ &$5.67$ & $5.23$&  $6.54$& $7.85$\\ 
     NS comp. & NS & SQS & NS & NS  & SQS & NS& NS \\
    \multirow{2}{*}{Newborn NS}     & NS & SQS: $j_{\rm ns,0}<1.25$ &  NS & NS  & SQS: $j_{\rm ns,0}<1.05$ & SQS: $j_{\rm ns,0}<1.05$&  NS \\ 
          & & NS: $j_{\rm ns,0}>1.25$ &   &  &SQS: $j_{\rm ns,0}>1.05$ & NS: $j_{\rm ns,0}>1.05$&  \\ 
          \hline 
      \multicolumn{8}{c}{\textbf{Binary outcome}} \\
      \hline
     & NS-NS  &  SQS-SQS   & NS-NS & NS-NS &  SQS-SQS &   NS-SQS &  NS-NS    \\
       &   &   SQS-NS &  &   & &  NS-NS &  \\
       \hline \hline
    \end{tabular}
    \end{adjustbox}
        \caption{Summary of the initial binary and the system fates for selected numerical simulations presented in this work.}\label{tab:outcome}
\end{table}

\section{Conclusions}
\label{sec:conclusions}

In this work, we have explored the possibility that quark deconfinement occurs during the process of massive accretion in the last stages of evolution of short-period binaries, at the moment of the second SN explosion experienced by the binary system. In particular, we have performed SPH numerical simulations of the SN explosion of a CO star in the presence of an NS companion in a compact orbit of a few-minute orbital period. The expelled matter by the SN accretes both onto the NS companion and by fallback onto the newborn NS formed from the gravitational collapse of the iron core of the CO star that led to the SN explosion.

For this task, we have monitored whether, during the accretion process, any of the NSs in the system reach the condition for SQM deconfinement, which we define as the strangeness fraction reaching the critical value $Y_S \approx 0.2$--$0.3$. We have explored representative values of the SN explosion (kinetic) energy and binary parameters such as the mass of the CO star. For the performed suite of simulations, we have found that both the NS companion and the newborn NS can reach the condition for conversion into SQS for some of the parameters explored. In particular, the conversion is favored for lower values of the SN kinetic energy and the initial angular momentum of the NS, i.e., before accretion begins. Furthermore, the simulations indicate that the newborn NS is most favored to reach the conversion condition.

We have shown that the conversion process of an NS into an SQS would release an enormous amount of energy, equivalent to a few $10^{52}$--$10^{53}$ erg, which suggests that the conversion process could produce highly energetic transients. The determination of the speed of the conversion of the entire NS into a SQS, as well as its associated observables need, however, calculations via hydrodynamics simulations, including radiative transfer and neutrino physics (see, e.g., \cite{2015PhRvC..92d5801D,2018PhLB..777..184O,2022PhRvC.105f5807K}, and references therein), which are beyond the scope of the present work and deserve forthcoming studies. 

This first suite of simulations suggests the possibility that these binaries may be progenitors of compact-object binaries with SQS components, if the mass loss from the SN explosion or the energy release in the NS conversion process does not unbind the binary. The merger time of such compact-object binaries is $\sim 10$ kyr. The outcome binary parameters found in this work can be helpful to investigate the qualitative and quantitative differences in the electromagnetic and gravitational-wave emission of SQS-NS or SQS-SQS mergers and SQS-SQS mergers relative to NS-NS mergers \cite{DePietri:2019khb,Bauswein:2015vxa,Prakash:2021wpz}.




\bibliographystyle{apsrev}
\bibliography{reference}

\end{document}